\documentclass{article}

\usepackage[utf8]{inputenc}
\usepackage{amsmath}
\usepackage{caption}
\usepackage{subcaption}
\usepackage{graphicx}
\usepackage{xcolor}
\usepackage{hyperref}
\usepackage{siunitx}

\usepackage{cite}

\usepackage{authblk}
\date{}
\title{Photonic reservoir computer based on frequency multiplexing}

\author[1]{Lorenz Butschek}
\author[1]{Akram Akrout}
\author[1]{Evangelia Dimitriadou}
\author[1, *]{Alessandro Lupo}
\author[2]{Marc Haelterman}
\author[1, +]{Serge Massar}

\affil[1]{Laboratoire d'Information Quantique, CP 224, Universit\'{e} libre de Bruxelles, Av. F. D. Roosevelt 50, B-1050, Bruxelles, Belgium}
\affil[2]{OPERA-Photonique, CP 194/5, Universit\'{e} libre de Bruxelles, Av. F. D. Roosevelt 50, B-1050, Bruxelles, Belgium}

\affil[*]{alessandro.lupo@ulb.be}

\affil[+]{serge.massar@ulb.be}

\begin{document}
\maketitle
\begin{abstract}
Reservoir computing is a brain inspired approach for information processing, well suited to analogue implementations.
We report a photonic implementation of a reservoir computer that exploits frequency domain multiplexing to encode neuron states.
The system processes 25 comb lines simultaneously (i.e. 25 neurons), at a rate of 20 MHz. We illustrate performances on two standard benchmark tasks: channel equalization and time series forecasting.
We also demonstrate that  frequency multiplexing allows  output weights to be implemented in the optical domain, through optical attenuation. We discuss the perspectives for high speed high performance low footprint implementations.

\end{abstract}

\section{Introduction}

The past decade has seen remarkable developments in the field of photonic neuro-inspired information processing, inspired by the dramatic successes of artificial intelligence and machine learning \cite{shen2017deep,zhang2021optical, lin2018all, feldmann2019all, xu202111}.
Among these neuro-inspired approaches, Reservoir Computing (RC) \cite{J01,MNM02,JH04,LJ09,LJS12} has been extensively studied, as it performs remarkably well for time-dependent signal processing, has fast training times, and is  simple to implement experimentaly. The first photonic RC approaches were based on time multiplexing of neurons \cite{P12,L12, A11}, which simplifies the implementation. Even though high speed implementation of this approach is possible\cite{B13, L17}, time multiplexing still implies an inherent slow down. Alternative approaches mainly consist in spatial multiplexing \cite{Bueno18, rafayelyan2020large, paudel2020classification,sunada2020using}, including integrated passive networks \cite{V14}.

The frequency degree of freedom of light allows for a large number of modes to be processed simultaneously: for instance, the telecom $C$ band spans \SI{4.4}{THz} and can therefore accommodate 220 channels spaced by \SI{20}{GHz}. Moreover, reliable and efficient optical components are commercially available to generate and manipulate the frequency degree of freedom. Recently these advantages have been exploited to realise perceptrons \cite{xu2020photonic} and convolutional engines \cite{xu202111,feldmann2021parallel}, achieving very high speeds.

Here we demonstrate a photonic RC in which the reservoir neuron states are encoded in the complex amplitudes of different lines of a frequency comb. Physically, the reservoir consists of an optical fiber cavity in which the comb lines are mixed by an electro-optic phase modulator driven by a radio frequency signal (following ideas used in quantum optics \cite{Bloch07,Olislager10}). 
The reservoir evolution is linear, with a quadratic nonlinearity introduced during readout by the photodiode, as introduced in \cite{V14,V15}.

The present work constitutes an important improvement of previous unpublished \cite{Akrout16, butschek2020parallel} and published \cite{Butschek19FrequencyMultiplexed} reports on the same system. The improvement concerns number of neurons, stability, overall performances and implementation of optical output weights.
Numerical analysis of an analog output layer and of an integrated setup were presented in \cite{Akrout2017autonomous} and \cite{Kassa18integrated} respectively.  The combination of the system presented here with time multiplexing was studied, in simulation, in \cite{ZippStoker21}. The present experimental system can be modified into an Extreme Learning Machine (ELM) \cite{Lupo2021}. We recommend that the reader read in parallel the ELM realisation as the principle of operation is simpler (there is no recurrence), but many of the technical aspects are identical.

A RC \cite{J01,MNM02,JH04} is a recurrent neural network exhibiting memory about past inputs, typically used to process time series. It is a randomized neural network, meaning that most of the weights are kept fixed and only the output weights are trained, avoiding the requirement for time and power expensive training algorithms  and simplifying the physical implementability. We implemented a linear reservoir, described by the equations
\begin{align}
  \mathbf{x}(n) &= \mathbf{W}\cdot\mathbf{x}(n-1)+\mathbf{W}^{in}\cdot f^{in}(u(n)), \label{eq:linear_reservoir}\\
  \mathbf{I}^{out}(n)&=f^{out}(\mathbf{W}\cdot\mathbf{x}(n-1)), \label{eq:output_variables}\\
  y(n)&=\mathbf{W}^{out}\cdot\mathbf{I}^{out}(n), \label{eq:output}
\end{align} 
where $\mathbf{x}(n)$ is the vector representing the state of the reservoir at the timestep $n$, $\mathbf{W}$ is the set of internal connection weights, $\mathbf{W}^{in}$ the set of input weights, $\mathbf{I}^{out}(n)$ is the vector representing the set of reservoir output variables at timestep $n$, $f^{in}$ and $f^{out}$ are the input and output nonlinearity respectively, $u(n)$ is the input time series, and $\mathbf{W}^{out}$ is the set of output weights optimized so that the reservoir output $y(n)$ is as similar as possible to the target. In software implementations, $\mathbf{W}$ and $\mathbf{W}^{in}$ are selected at random and kept fixed while $\mathbf{W}^{out}$ is trained. In physical implementations, $\mathbf{W}$ and $\mathbf{W}^{in}$ are usually determined by the chosen physical system. See appendix \ref{appendix:model} for more details on the model describing $\mathbf{W}$ and $\mathbf{W}^\textrm{in}$.

%% Setup
\section{Experimental system}

\begin{figure}[b!]
    \centering
    \includegraphics[width=0.95\textwidth]{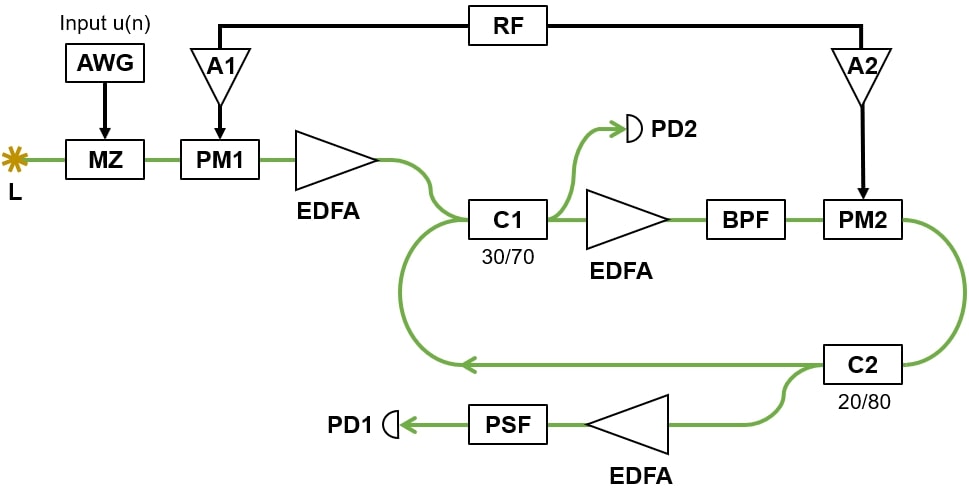}
    \caption{Schematic of the experiment. Black lines: electrical connections. Green lines: polarization maintaining optical fiber. L: laser. AWG: Arbitrary Waveform Generator. MZ: Mach Zehnder modulator. PM1 and PM2: Phase Modulators. EDFA: Erbium Doped Fiber Amplifier. A1 and A2: RF Amplifiers. C1 and C2: Couplers. BPF: Band Pass Filter. PSF: Programmable Spectral Filter. PD1 and PD2: Photodiodes.}
    \label{fig:setup}
\end{figure}

\begin{figure}[t]
    \centering
    \includegraphics[width=0.55\textwidth]{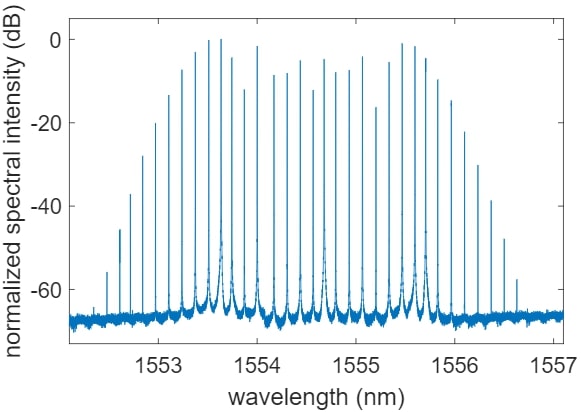}
    \caption{Frequency comb measured at the output of PM1 with the input $u(t)=0$ set to zero. (Center wavelength: $\lambda=\SI{1554.6}{nm}$; RF modulation frequency: $\Omega/2\pi=\SI{16.99175}{GHz}$.)}
    \label{fig:comb}
\end{figure}

The setup depicted in Figure \ref{fig:setup} is entirely built from single-mode polarisation-maintaining optical fiber.
% and consists in three sections which we call "input layer", "reservoir" and "output layer".
The input layer contains a $C$-band CW narrow band laser whose wavelength $\lambda= 2 \pi c/\omega$ can be tuned in the range \SIrange{1554.5}{1555.5}{nm}. An Arbitrary Waveform Generator (AWG) encodes the input $u(n)$ in the laser radiation through a Mach-Zehnder modulator (MZ). The input $u(n)$ is held constant for a time equal to the round trip time of the cavity.
The MZ defines the input nonlinearity $f^{in}$ in \eqref{eq:linear_reservoir}: $f^{in}(u(n))=E_0 \sin (\gamma u(n) + \pi/4)$, where $\gamma$ depends on the amplitude of the signal generated by the AWG and $E_0$ is the amplitude of the laser radiation leaving the MZ modulator. The monochromatic radiation encoding $u(n)$ then passes through a Phase Modulator (PM1) driven by a periodic RF signal oscillating at frequency $\Omega$ and amplified by Amplifier A1, thereby generating a frequency comb, see Figure \ref{fig:comb}. An EDFA increases the optical power to approximately \SI{17}{dBm} before injecting the signal into the reservoir.

The reservoir layer consists in an optical fiber loop with roundtrip time $\tau = \SI{49.5}{ns}$, corresponding to a Free Spectral Range of \SI{20.2}{MHz}. 
The cavity contains a second EDFA, 
%(which we define "intra-cavity"), 
a \SI{3}{nm} wide Band-Pass Filter (BPF)  to suppress amplified spontaneous emission noise, and a second Phase Modulator, PM2. PM2 is driven by the same RF signal driving PM1, but through  Amplifier A2. Since PM2 acts on radiation already featuring a comb-like spectrum,  the phase modulation results in line interference. 
Thus at any point in the cavity (say for definiteness just after C2) the electric field amplitude can be represented (during roundtrip $n$, at time $t$) as
\begin{equation}
    E_n(t)=\sum_k x_k(n)e^{-i(\omega+k\Omega)t}.
    \label{Eq:En(t)}
\end{equation}
The  collection of comb lines amplitudes $\{x_k(n)\}$ in \eqref{Eq:En(t)}  should thus be identified with the
vector $\mathbf{x}(n)$ in Eqs. (\ref{eq:linear_reservoir}, \ref{eq:output_variables}).

\begin{figure}[t]
         \centering
         \includegraphics[width=0.55\textwidth]{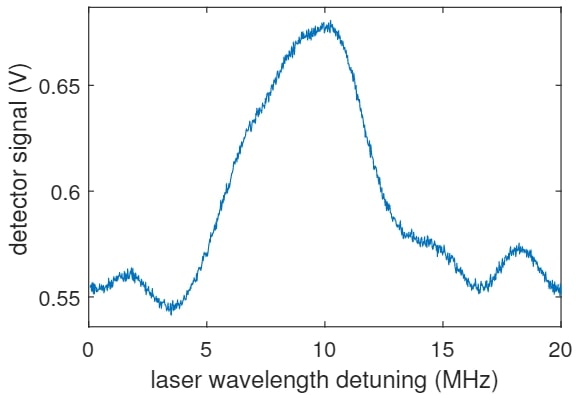}
         \caption{Example of transfer function of the experimental setup as a function of the laser wavelength shift. The vertical axis is the output of photodiode PD2 when the input $u(t)=0$ is set to zero. (Center wavelength: $\lambda=\SI{1554.6}{nm}$; RF modulation frequency: $\Omega/2\pi=\SI{16.99175}{GHz}$.) Visualization 1 contains a video showing the strong dependence of the transfer function on $\Omega$.}
         \label{fig:transfer_function}
\end{figure}

The cavity constitutes a complex  interferometer in the frequency domain, see Figure \ref{fig:transfer_function} and appendices \ref{appendix:omega} and \ref{appendix:model}. The  stabilization mechanism of the cavity, not shown in Figure \ref{fig:setup}, consists in a PID controller reading the average cavity power through PD2 and both driving a Peltier cell (against thermal drifts) and piezo-tuning the laser wavelength (against acoustical noises). 
The piezo-tuning allows adjustment of the laser wavelength of approximately \SI{1}{pm} at frequencies below \SI{10}{kHz}. 
The optical cavity, including the intra-cavity EDFA, is mounted in an insulated box.

Part of the radiation leaves the optical cavity through Coupler C2 and reaches the output layer. Here the radiation is amplified by an EDFA, reaching an optical power of \SI{10}{dBm}, filtered by a Programmable Spectral Filter (PSF -- Finisar Waveshaper) and read by the photodiode PD1. The PSF either applies output weights $\mathbf{W}^{out}$ (i.e.\ optical attenuation) on all comb lines simultaneously, or is used to read each comb line (i.e.\ each neuron) independently by implementing a notch filter. Note that comb states are encoded in complex amplitudes, while PD1 reads intensities, resulting in a quadratic  output function: $f^{out}(\mathbf{x}(n))=|\mathbf{x}(n)|^2$ (with the norm square $|\cdot|^2$ acting elementwise).

Key components in our experiment are the phase modulators PM1 and PM2.
An ideal phase modulator driven by a periodic electric field oscillating at frequency $\Omega$ acts on a monochromatic CW laser radiation at frequency $\omega$ according to
\begin{eqnarray}
E_0 e^{-i\omega t} &\rightarrow& E_0 e^{-i \omega t} e^{-i m \cos (\Omega t)}
\nonumber\\ & &
= E_0 e^{-i \omega t} \sum_k i^k J_k(m) e^{-i k \Omega t}.
\label{eq:pm_sidebands}
\end{eqnarray}
 where $J_k$ is the $k$-th order Bessel functions of the first kind, and  $m=\pi V/V_\pi$ is the modulation index, with $V$ the RF signal amplitude and $V_\pi$ the characteristic voltage of the phase modulator. The second line of \eqref{eq:pm_sidebands} is known as the Jacobi-Anger expansion.
\eqref{eq:pm_sidebands} shows that periodic phase modulation of monochromatic radiation generates a frequency comb whose line are spaced by $\Omega$ (see Figure \ref{fig:comb}). Greater $m$ coefficients result in broader combs. \eqref{eq:pm_sidebands}  also implies that if the input of the PM is a frequency comb with lines spaced by $\Omega$, the output of the PM will still be a frequency comb with lines spaced by $\Omega$, but with interferences between the line amplitudes.

In our experiment PM1 and PM2 are driven  by  amplifiers A1 and A2 whose output powers are \SI{32}{dBm} and \SI{20}{dBm} resulting in modulation indexes
$m_1=7.9$ and  $m_2=2.2$ respectively.
The RF frequency $\Omega/2\pi$, defining the spacing of the comb, is selected in the range $15-\SI{18}{GHz}$ and kept constant for the whole duration of the experiment. 
The input weights $W^{in}$ in \eqref{eq:linear_reservoir} are thus determined by the modulation index $m_1$ of PM1. The interconnection matrix $\mathbf{W}$ in \eqref{eq:linear_reservoir} is determined by the modulation index $m_2$ of PM2, by the roundtrip phase accumulated by each line of the comb (which differs for each line due to the group velocity), and by the overall amplitude attenuation $\alpha$ of the cavity.

Experimentally we measure a small deviation from \eqref{eq:pm_sidebands} which can be well fitted by assuming a higher harmonic contribution to the phase modulation (see \cite{Lupo2021} for details). Numerical simulations suggest this contribution of higher harmonics is beneficial to the performance of the reservoir, presumably because it results in asymmetries both in the input weights $\mathbf{W}^{in}$ and in the  matrix $\mathbf{W}$.

The setup  allows for two readout modes.
In the first  "digital"  readout mode, the experiment is executed as many time as there are the comb lines to read. During each execution  the same input $u(n)$ is supplied and the reservoir evolution is  the same. During each run, a bandpass filter selecting only one comb line is set on PSF. Thus, PD1 records each time a different component of $\mathbf{I}^{out}(n)$. 
After all the components of $\mathbf{I}^{out}(n)$ 
are recorded, the optimal output weights $\mathbf{W}^{out}$ are computed and the multiplication by $\mathbf{W}^{out}$ in \eqref{eq:output} is carried out on the computer.

The second readout mode uses "optical weighting", as proposed in \cite{Akrout2017autonomous} and demonstrated previously in \cite{Lupo2021}.
To implement "optical weighting", the output weights computed in "digital" output mode are divided in  two sets, corresponding to the positive and negative weights.
The experiment is then executed twice, injecting the same input $u(n)$. During the first run  attenuations proportional to the positive weights are set on PSF and the attenuated signal $y^+(n)$ is recorded by PD1; during the second run attenuations proportional to the negative weights are set on PSF and the attenuated signal $y^-(n)$ is recorded by PD1. Weights are normalized such that attenuations always span the range \SIrange{-60}{0}{dBm}. We then take the output to be 
\begin{equation}
y(n)=C^+ y^+(n) - C^- y^-(n)+C^0
\label{Eq:opticalweights}
\end{equation}
where the three parameters $(C^+,\ C^-,\ C^0)$ are optimised, and the computation of $y(n)$ is carried out on the computer.

\begin{figure*}
     \centering
         \includegraphics[width=1\textwidth]{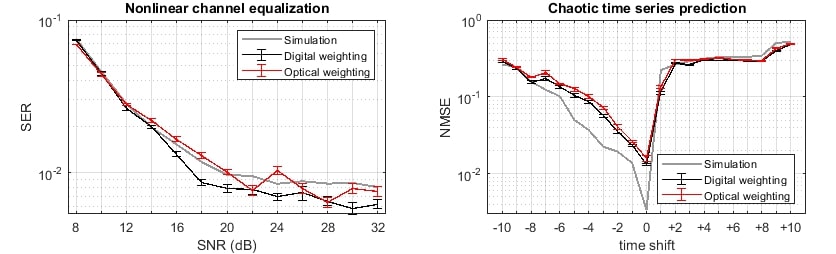}
         \caption{Benchmark results. Error bars represent the standard deviation of 10 different runs of the experiment with different data. Note that digital weighting and optical weighting give comparable results. The numerical simulation accounts for phase noise and detector noise.}
         \label{fig:plots}
\end{figure*}

When testing our RC on benchmark tasks, the AWG amplitude is set such that $\sin^2(\gamma u(t)+\pi/4) \in [0.28,0.72]$, and 
the overall amplitude attenuation of the cavity is taken to be  $\alpha=0.70$.
The choice of $\Omega$ sensibly affects the cavity transfer function (Fig. \ref{fig:transfer_function}) and the reservoir performances (see appendix \ref{appendix:omega}), hence this parameter is optimized for each task. The modulation frequency $\Omega$ is the only parameter differing among the two tasks described below. For the readout layer we use  only the 25 central lines of the frequency comb, as the other lines are too weak to be useful. The optimisation of the readout weights $\mathbf{W}^{out}$ is done by minimising the Mean Square Error (MSE) and using Ridge regression.

\section{Results}

We report here results on two benchmark tasks: nonlinear channel equalization, following the model reported in \cite{JH04}, and a time-series prediction of the recorded output of a chaotically operating far-infrared laser (Santa-Fe)\cite{SANTAFE93} (results on other tasks can be found in \cite{Akrout16}). For both tasks, digital weighting and optical weighting of the outputs give comparable results.

In the first task (Fig. \ref{fig:plots}, left panel), symbols randomly selected from $\{-3,\ -1,\allowbreak \ +1,\ +3\}$ are propagated through a simulated channel exhibiting nonlinearity, memory and noise. Zero-mean Gaussian noise is adjusted in power to obtain a Signal-to-Noise Ratio (SNR) ranging from \SIrange{8}{32}{dB}. The signal $u(n)$ received at the end of the communication channel is used as the input of the RC, trained to recover the original signal. Performances are evaluated by calculating the Symbol Error Rate (SER), defined by the fraction of the misclassified symbols within the generated sequence. After a first sequence of 10000 symbols used to remove transients, the reservoir is trained on a sequence of 5000 symbols, and then tested on 5000 symbols. Our RC performs comparably to previous implementations \cite{P12, V15, D12, Dejonckheere:14, Lupo2021} for SNRs up to \SI{16}{dB}: we reach SER values of $0.0166\pm0.0007$ and $0.0133\pm0.0005$ for optical and digital weighting respectively at SNR of \SI{16}{dB}. For higher SNR values performance saturates because of system noise.
% Dig mean
% 0.0062    0.0059    0.0065    0.0074    0.0070    0.0077    0.0079    0.0086    0.0133    0.0200    0.0265    0.0449    0.0740
% Dig std
% 0.0004    0.0005    0.0004    0.0008    0.0003    0.0005    0.0004    0.0004    0.0005    0.0004    0.0011    0.0014    0.0007
% Opt mean
% 0.0076    0.0079    0.0064    0.0079    0.0103    0.0076    0.0101    0.0130    0.0166    0.0219    0.0280    0.0439    0.0692
% Opt std
% 0.0005    0.0006    0.0005    0.0006    0.0006    0.0006    0.0004    0.0006    0.0007    0.0006    0.0006    0.0011    0.0004

In the second task (Fig. \ref{fig:plots}, right panel), a timeseries is supplied to the reservoir, which is asked to predict future evolution or to recall past inputs. The first 200 steps of the dataset are discarded as a warmup sequence, after which the training sequence consists of 2670 steps, followed by 2470 steps for testing. The one-step-ahead prediction performances are comparable with previous works \cite{A11, L12}: normalized mean square error (NMSE) values are $0.134\pm0.005$ and $0.113\pm0.005$ for optical and digital weighting respectively. 

Fig. \ref{fig:plots} also contains results from a numerical simulation accounting both for detector noise ($\textrm{SNR}=24\ \textrm{dB}$) and phase noise (gaussian with $\sigma = 16\ \textrm{mrad}$). Details are provided in appendix \ref{appendix:sim}.  

\section{Conclusion}

The present experiment lends itself to several improvements. The use of a mode locked laser as optical source could provide a much broader frequency comb at the input, resulting in a reservoir with much more neurons.  The last step in optical weighting mode \eqref{Eq:opticalweights} could readily be implemented using a balanced photodectector, resulting in a completely analogue output layer. The processing speed of the experiment could be significantly increased by using a smaller cavity, or by combining frequency and time multiplexing as proposed in \cite{ZippStoker21}.

One of the main perspectives of the present experiment is the development of an integrated version. A preliminary study \cite{Kassa18integrated} suggests that it should be possible to integrate on an InP chip, with a RF modulation frequency of $\Omega/2\pi=\SI{10}{GHz}$ and a much shorter cavity with a free spectral range of \SI{2.5}{GHz}. This would both simplify stabilization and considerably increase data throughput. 

In conclusion the present work shows how reservoir computing, a brain inspired approach to computation, can be implemented with light using frequency multiplexing. This offers a new route to developing compact, high performance, optical information processing. 

\section*{Funding} The authors acknowledge financial support by the FRS-FNRS grants PDR T.0092.14, CDR J.0040.16, PDR T.0089.18, CDR J.0130.21, and by the European Commission grant 860360–POSTDIGITAL.

\section*{Acknowledgments} The authors thank Q. Vinckier and A. Bouwens for their contribution to an early version of this experiment.
%\bmsection{Disclosures} The authors declare no conflicts of interest.

%\section*{Data availability} Data underlying the results presented in this paper are not publicly available at this time but may be obtained from the authors upon reasonable request.
\appendix
\section{Dependence of performances on phase modulation frequency}
\label{appendix:omega}
The optical cavity is a complex  interferometer whose behaviour strongly depends on the frequencies of the circulating radiation. The two parameters defining these frequencies are the laser frequency $\omega$, which determines the position of the central comb line, and the RF phase modulation frequency $\Omega$, which is the spacing of the comb lines.

A rough characterisation of the optical cavity is its transfer function which we measure by recording optical power reflected by the cavity onto photodiode PD2 when the laser frequency $\omega$ is varied. In a typical transfer function measurement we scan $\omega$ in a $20\textrm{ MHz}$ range around its central value. $20\textrm{ MHz}$ is the experimentally accessible range closest to the cavity free spectral range, which is the period of the transfer function. The video reported in Visualization 1 shows how the cavity transfer function changes its shape when $\Omega$ is shifted: for each value of $\Omega$  we scan $\omega$ in a $20\textrm{ MHz}$ range.

\begin{figure}
    \centering
    \includegraphics[width=\textwidth]{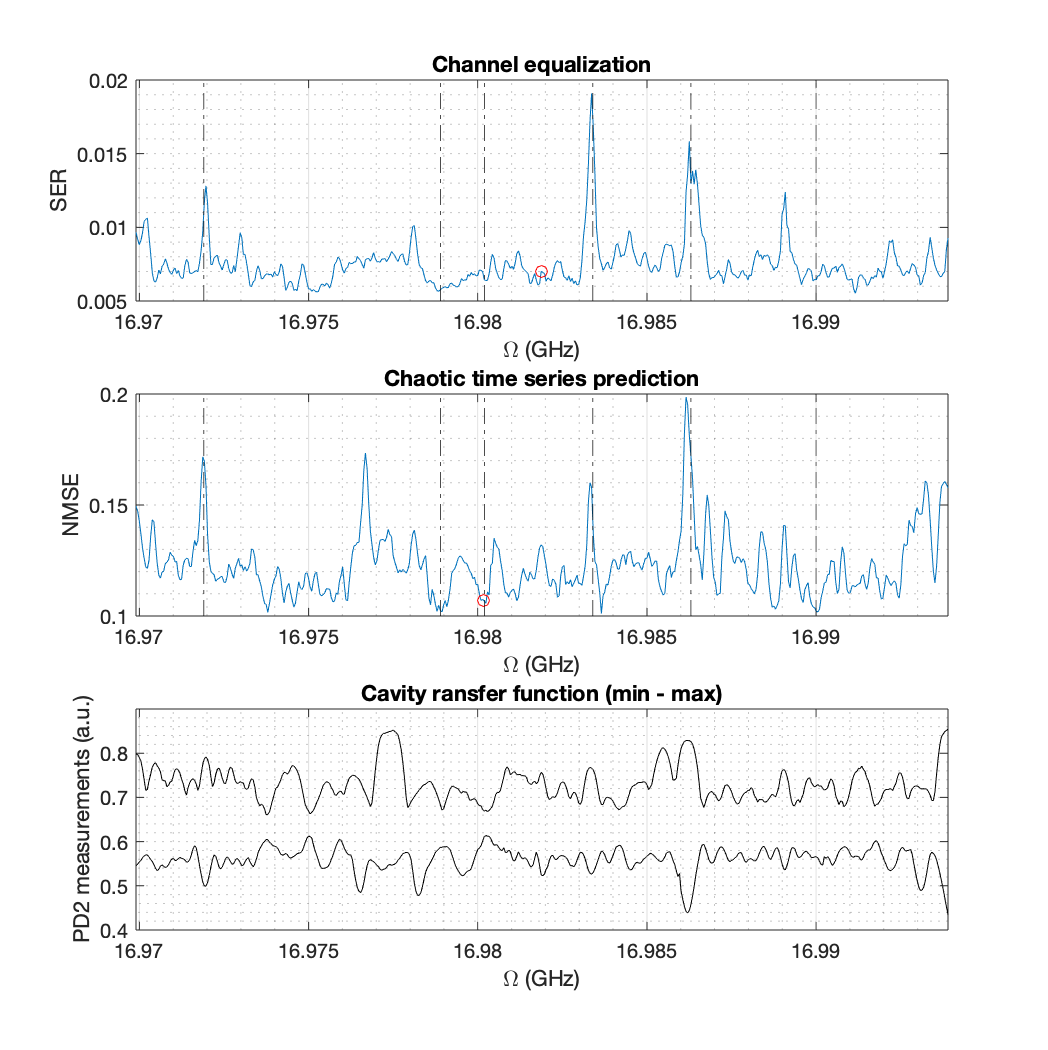}
    \caption{(top and middle) Performance on nonlinear channel equalization task (top panel, SNR is set on $32$ dB) and chaotic time series prediction (middle panel, time-shift is set on $+1$) varying $\Omega$. The red circle indicates the configuration in which we run the experiment reported in the main text. The laser wavelength, defining the position of the central comb line, is the same employed for the experiment reported in the main text ($\lambda=1554.6\textrm{ nm}$). The vertical lines indicate the configurations which we analyzed further in Fig. \ref{fig:tfs}. (bottom) Summary representation of the transfer function. Only minimum and maximum values are reported. 
    }
    \label{fig:omega_sweep}
\end{figure}

The behaviour of the cavity determines the way in which comb lines interfere with each other and thus determines the set $\mathbf{W}$ of internal connection between the neurons of the reservoir (see Eq.\ (1) in the main text). 
The Reservoir Computing scheme relies on random internal connections, but in experimental systems such as the present one, in which only a few parameters can be tuned, some parameters result in better performance than others. In particular 
certain sets of parameters may give rise to a bad operating condition.
In Fig.\ \ref{fig:omega_sweep} we plot the experimentally measured performances on both the tasks reported in the main text while sweeping $\Omega$ in the range between $16.970$ GHz and $16.994$ GHz. The figure also reports a compressed representation of the cavity transfer function where only minimum and maximum values are plotted. We do not see a clear correlation between the transfer function extension and the reservoir performances. 

\begin{figure}
     \centering
     \begin{subfigure}[b]{0.45\textwidth}
         \centering
         \includegraphics[width=\textwidth]{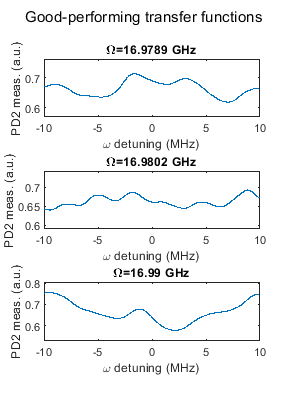}
         \caption{}
         \label{fig:y equals x}
     \end{subfigure}
     \hfill
     \begin{subfigure}[b]{0.45\textwidth}
         \centering
         \includegraphics[width=\textwidth]{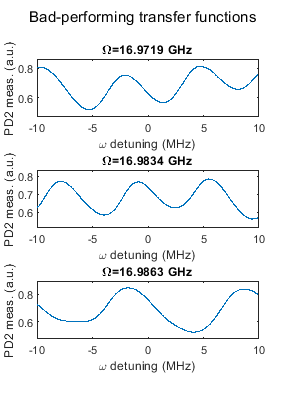}
         \caption{}
         \label{fig:three sin x}
     \end{subfigure}
     \caption{Example of cavity transfer function measured for values of $\Omega$  giving good performances (a) and bad performances (b). }
     \label{fig:tfs}
\end{figure}

Fig.\ \ref{fig:tfs} reports some examples of transfer function measured both for good-performing $\Omega$ values and bad-performing ones. We do not identify a clear relation between transfer function shapes and performances, but it appears  that  bad performing configurations are characterized by transfer functions less complex than usual and almost sinusoidal, which could be related to the presence of a resonance in the cavity. Note that most of the $\Omega$ values giving bad performances are common to both tasks.

\section{Analytical Model}
\label{appendix:model}

\subsection{Introduction}
Here we describe the model behind the numerical simulation whose results are reported in the main text (Fig.\ 4 in the main text). The model runs in discrete time (a timestep is a roundtrip around the cavity) and accounts for noise in the photodiode measurements, for phase noise in the cavity and for refractive index dispersion (different wavelengths see different refractive indexes and accumulate different phases during the propagation). When set with realistic values for detector noise, phase noise and index dispersion, the simulation is in good agreement with experimental data. However in its current state it does not reproduce the detailed dependence of the performances on $\Omega$ (see experimental data reported in Fig.\ \ref{fig:omega_sweep}), most likely because we do not know the exact length of optical fiber between optical components.

We represent the complex field amplitude of each of the N comb lines as an element of a $N\times 1$ vector, such that the central element of the vector is the amplitude of the central comb line. The state of the system is then defined in each timestep $n$ by two vectors: $\mathbf{x}^\textrm{in}(n)$ and $\mathbf{x}(n)$, representing respectively the amplitudes of the lines of the input comb and the amplitudes of the lines of the comb propagating inside the cavity. Both vectors are assumed to represent the field at the entrance of the cavity, at the coupler $C1$ (see Fig.\ 1 in the main text).

\subsection{Phase modulation}
The effect of the first phase modulator, PM1, which is placed before the cavity, is represented by the $N\times 1$ complex vector $\mathbf{W}^\textrm{in}$ such that $\mathbf{W}^\textrm{in}_j$ represents how strongly the input radiation is coupled with the $j$-th comb line, i.e.\ the $j$-th neuron. The effect of the second phase modulator, PM2, which is placed inside the cavity, is represented by the $N\times N$ complex matrix $\mathbf{W}^\textrm{PM}$ such that $\mathbf{W}^\textrm{PM}_{j,\ k}$ represents how strongly the phase modulation couples the $j$-th comb line (i.e.\ the $j$-th neuron) with the $k$-th one.

\begin{figure}
     \centering
     \begin{subfigure}[b]{0.49\textwidth}
         \centering
         \includegraphics[width=\textwidth]{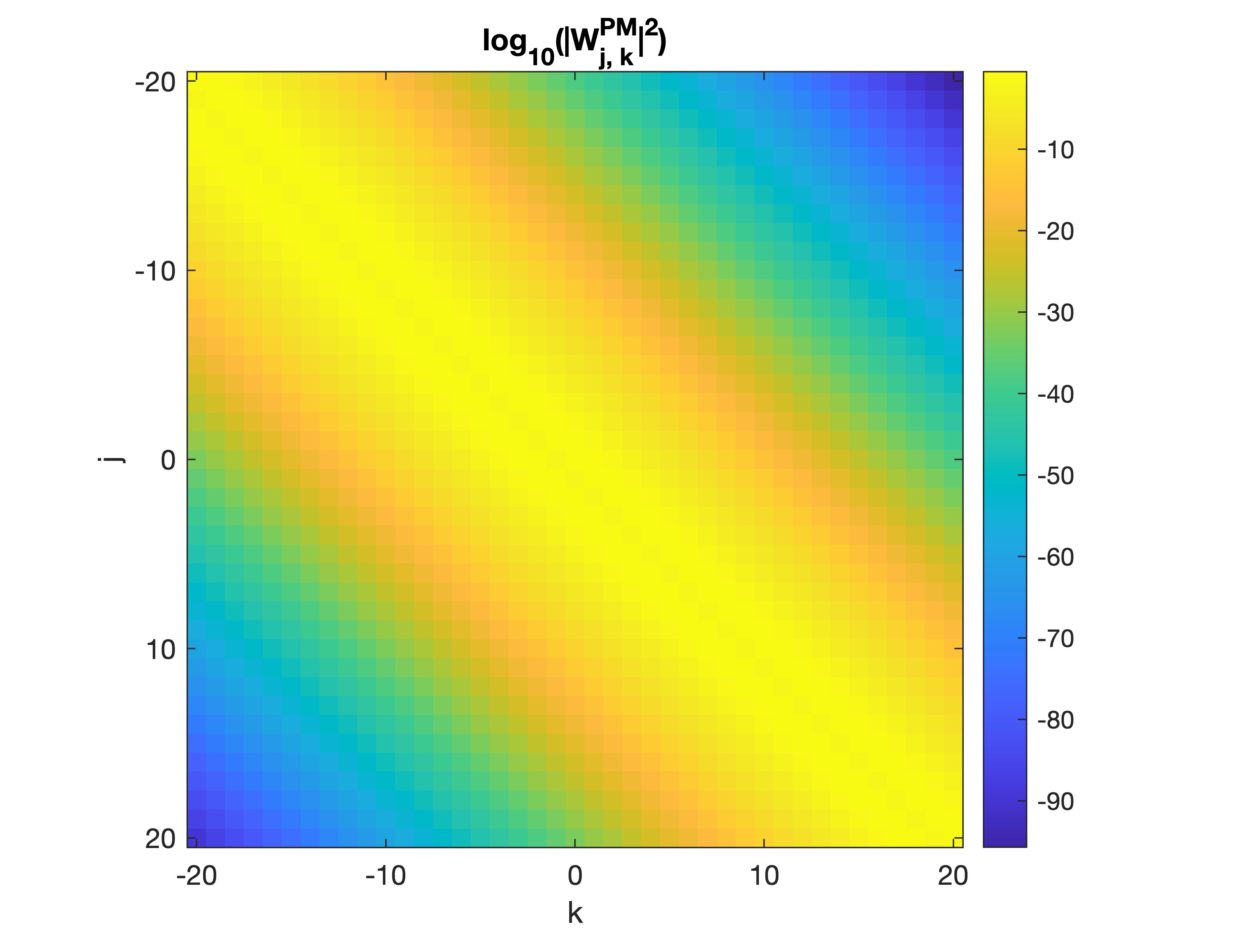}
         \caption{}
     \end{subfigure}
     \hfill
     \begin{subfigure}[b]{0.49\textwidth}
         \centering
         \includegraphics[width=\textwidth]{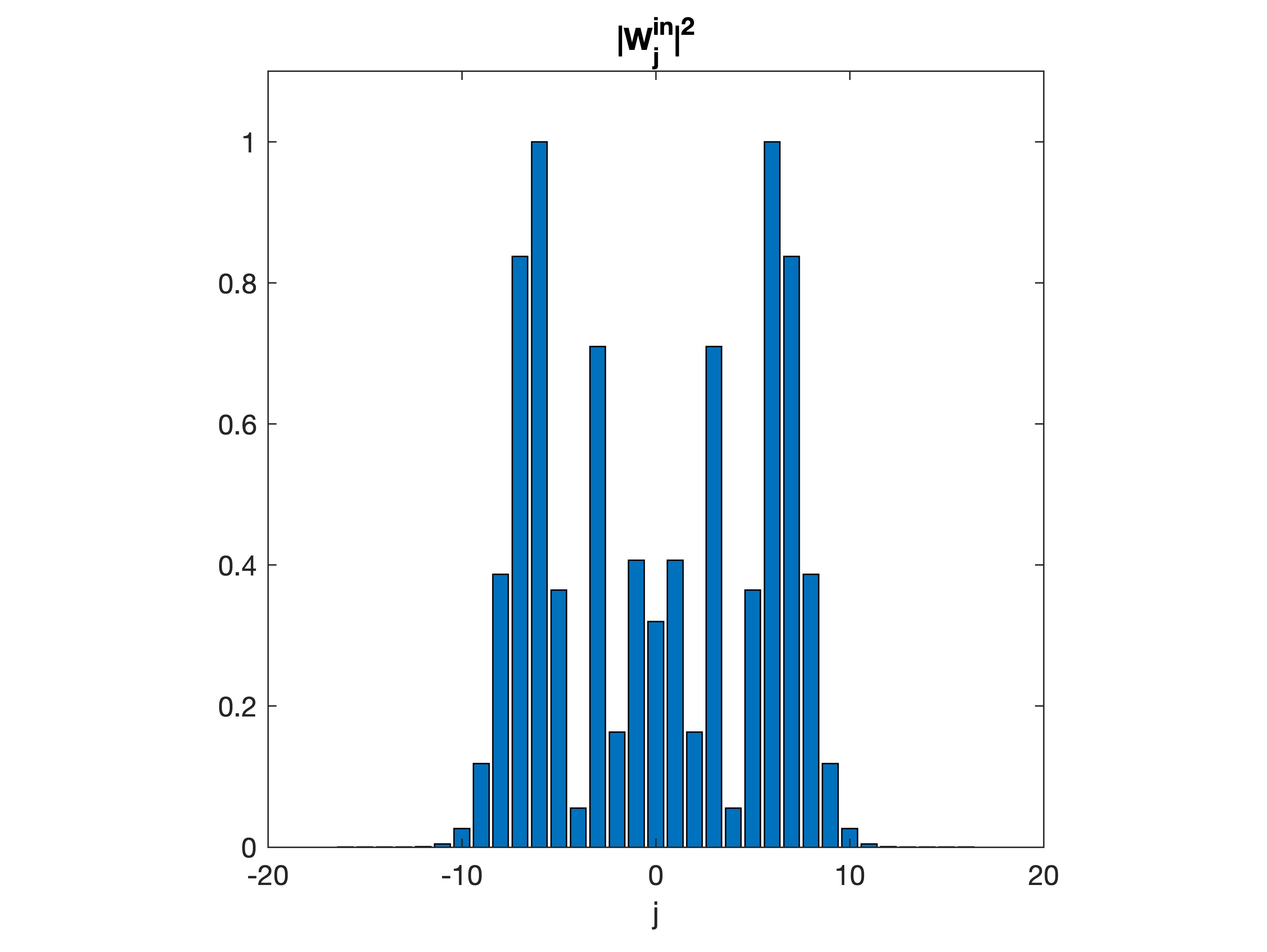}
         \caption{}
     \end{subfigure}
     \caption{Graphical representations of the matrix $\mathbf{W}^\textrm{PM}$ (a) and the vector $\mathbf{W}^\textrm{in}$ (b) (both normalized) defining respectively the set of connections between neurons and the set of input weights.}
     \label{fig:matrices}
\end{figure}

$\mathbf{W}^\textrm{in}$ and $\mathbf{W}^\textrm{PM}$ can be derived based on the effect described by Eq.\ (5) in the main text:
\begin{equation}
    \mathbf{W}^\textrm{in}_j = i^{j-N_0}J_{j-N_0}(m_1),
\end{equation}
\begin{equation}
   \mathbf{W}^\textrm{PM}_{j,\ k} = i^{j-k}J_{j-k}(m_2),
\end{equation}
where $N_0$ is the index of the $\mathbf{x}$ vector element representing the central comb line (if $N$ is odd, $N_0=(N+1)/2$), while $J_k$ is the Bessel function of the first kind and $m_1$ and $m_2$ are adimensional values representing the strength of the modulations generated by PM1 and PM2 respectively. Fig.\ \ref{fig:matrices} shows the shape of $\mathbf{W}^\textrm{in}$ and $\mathbf{W}^\textrm{PM}$.

\subsection{Input and propagation in the optical cavity}
At the timestep $n$ the Mach Zehnder modulator is driven by the signal $u(n)$ ($u(n)\in[-1,\ +1]$). We model the effect of the modulator on the light amplitude, $E^\textrm{in}(n)$, as
\begin{equation}
    E^\textrm{in}(n) = E_0\sin(\gamma u(n)+\frac{\pi}{4}),
\end{equation}
where $\gamma$ represents the modulator driving strength. In the simulation we assume that the light amplitude  input to the modulator is constant and unitary ($E_0=1$)  and we neglect the effect of the propagation outside the cavity, hence $E^\textrm{in}(n)$ is real. The comb that at timestep $n$ enters the cavity is described by the vector
\begin{equation}
    \mathbf{x}^\textrm{in}(n) = \beta\cdot\mathbf{W}^\textrm{in}\cdot E^\textrm{in}(n),
\end{equation}
where $\beta<1$ is a factor accounting for the coupling losses toward the cavity.

The effects of a roundtrip inside the cavity are summarized in the complex $N\times N$ matrix 
\begin{equation}
\mathbf{W}=\alpha\cdot\Phi^\textrm{noise}\cdot\mathbf{\Phi}^{(2)}\cdot\mathbf{W}^\textrm{PM}\cdot\mathbf{\Phi}^{(1)},
\label{Eq:S5}
\end{equation}
where $\alpha$ is a real scalar value accounting for losses (including the amount of power leaving the cavity for readout) and gain (optical amplifier) affecting the radiation in one roundtrip, $\Phi^\textrm{noise}$ is an unit-modulus scalar value randomly extracted at each roundtrip to represent phase noise, while $\mathbf{\Phi}^{(1)}$ and $\mathbf{\Phi}^{(2)}$ are complex diagonal matrices whose  
elements account for the phase accumulated by each comb line during propagation respectively between the cavity entrance and PM2 and between PM2 and the cavity entrance. For $l\in[1,\ 2]$:
\begin{equation}
    \mathbf{\Phi}^{(l)}_{j,\ k}=
    \begin{cases} 
    e^{2i\pi n(\omega_j)L_l} & \mbox{if } j=k \\ 
    0 & \mbox{otherwise}
    \end{cases},
    \label{Eq:S6}
\end{equation} 
where $n(\omega_j)$ is the refractive index seen by wavelength $\omega_j$ (wavelength of the $j$-th comb line) and $L_1$ and $L_2$ represent respectively the length of the path between the cavity entrance and PM2 and the length of the path between PM2 and the cavity entrance. 

For $n(\omega_j) $ we use the Sellmeier equation for single mode fiber. However,  linearizing $n(\omega_j)$ as
\begin{equation}
n(\omega_j) = \beta_0 + j  \Omega \beta_1
\label{Eq:varphi}
\end{equation}
show negligible differences, where $\beta_0$ is the  propagation constant at frequency $\omega$ and $\beta_1 = v_g^{-1}$  the inverse of the group velocity.

For the propagation lengths we use $L_1= p L$ and $L_2= (1-p)L$ with $L=10$ m, and $p \in [0,1]$. Simulations show that $p$ can be chosen arbitrarily, except for a few values for which performance decreases.

At the timestep $n$ the comb inside the cavity is given by
\begin{equation}
\label{eq:x}
    \mathbf{x}(n) = \mathbf{W}\cdot\mathbf{x}(n-1)+\mathbf{x}^\textrm{in}(n).
\end{equation}

\subsection{Absorbing propagation phases of the input}

Note that in \eqref{eq:x} we do not take into account the phases accumulated by the comb lines between PM1 and C1. However,  we now show this is not a loss of generality as we can absorb this phase in a redefinition of  $\mathbf{x}(n)$.
Indeed, denote the action of propagation between  PM1 and C1 by the matrix $\mathbf{\Phi}^{(3)}_{j,\ k}$ (similar to  \eqref{Eq:S6}). Then  \eqref{eq:x} should be replaced by
\begin{equation}
    \mathbf{x}(n) = \mathbf{W}\cdot\mathbf{x}(n-1)+\beta\cdot\mathbf{\Phi}^{(3)}\cdot \mathbf{W}^\textrm{in}\cdot E^\textrm{in}(n).
    \label{Eq:S9}
\end{equation}
If we define 
\begin{equation}
    \mathbf{x}(n) = \mathbf{\Phi}^{(3)} \cdot  \mathbf{x}'(n) 
    \label{Eq:S10}
\end{equation}
then  \eqref{Eq:S9} takes the form
\begin{equation}
    \mathbf{x}'(n) = \alpha\cdot\Phi^\textrm{noise}\cdot\mathbf{\Phi}'^{(2)}\cdot\mathbf{W}^\textrm{PM}\cdot\mathbf{\Phi}'^{(1)} \cdot\mathbf{x}'(n-1)+\beta\cdot \mathbf{W}^\textrm{in}\cdot E^\textrm{in}(n).
    \label{Eq:S11}
\end{equation}
where $\mathbf{\Phi}'^{(2)} = \mathbf{\Phi}^{(3)-1} \cdot \mathbf{\Phi}^{(2)}$ and $\mathbf{\Phi}'^{(1)} = \mathbf{\Phi}^{(1)} \cdot \mathbf{\Phi}^{(3)}$. This has exactly the same form as \eqref{eq:x}, but with different matrices $\mathbf{\Phi}^{(1)} $ and $\mathbf{\Phi}^{(2)}$.

\subsection{Readout}

In the numerical model we assume that the output coupler $C2$ is placed just before coupler $C1$. Consequently, at timestep $n$ the optical intensities of the output comb lines are described by the $N\times 1$ vector 
\begin{equation}
    \label{eq:iout}
    \mathbf{I}^\textrm{out}(n) = |\mathbf{W}\cdot\mathbf{x}(n-1)|^2,
\end{equation}
where the $|\cdot|^2$ operation acts element-wise. The multiplication by $\mathbf{W}$ takes into account that the radiation is extracted just before the injection of the new input at $C1$, in other words, an input always propagates at least once round the cavity before being extracted at the output.

We define $\mathbf{F}$ the $1\times N$ vector describing the attenuation that the spectral filter applies to each comb line, such that $\mathbf{F}_i$ is the attenuation applied to the $i$-th line. Hence, the power reaching PD1 is given by
\begin{equation}
    y(n) = \mathbf{F}\cdot\mathbf{I}^\textrm{out}(n) + y^\textrm{noise},
\end{equation}
where $y^\textrm{noise}$ is a scalar value randomly extracted at each timestep to simulate the detector noise.

\subsection{Analytical expression for output intensities}
%I have added this subsection to help one of the reviewers.

\eqref{eq:x} is a linear recurrence with source given by $E^\textrm{in}(n) $. 
If we neglect the phase noise $\Phi^\textrm{noise}$ in \eqref{Eq:S5}, then we can resum the recurrence to obtain 
\begin{equation}
 \mathbf{x}(n) = \sum_{k=0}^\infty  \beta\cdot \mathbf{W}^k \cdot \mathbf{W}^\textrm{in}\cdot E^\textrm{in}(n-k).
\end{equation}
Then using \eqref{eq:iout}, we see that the optical intensities of the output comb lines take the form
\begin{equation}
\mathbf{I}^\textrm{out}(n) = \lvert \sum_{k=0}^\infty  \beta\cdot \mathbf{W}^{k+1} \cdot \mathbf{W}^\textrm{in}\cdot E^\textrm{in}(n-k)\rvert^2 + y^\textrm{noise}
\end{equation}
which is a quadratic function of the previous inputs $(E^\textrm{in}(n), 
E^\textrm{in}(n-1),...)$.

\section{Simulation results}
\label{appendix:sim}
%I suggest to make a new section for the results.

In simulations we take the parameters to be $\alpha =0.754$, $\beta=0.43$, $\gamma=0.33$, $L_1=L_2=5\textrm{ m}$ , $\Omega=16.983\textrm{ GHz}$ and $\omega=(2\pi c)/\lambda$ where $\lambda=1554.6\textrm{ nm}$ and $c$ is the speed of light.

The simulations are in good agreement with the experimental results, as demonstrated by Fig.\ 4 in the main text.

\begin{figure}
     \centering
     \begin{subfigure}[b]{0.45\textwidth}
         \centering
         \includegraphics[width=\textwidth]{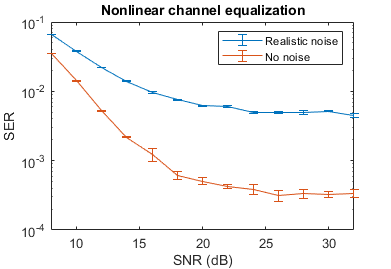}
         \caption{}
     \end{subfigure}
     \hfill
     \begin{subfigure}[b]{0.45\textwidth}
         \centering
         \includegraphics[width=\textwidth]{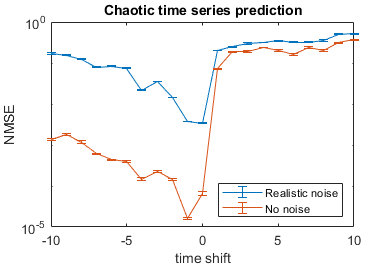}
         \caption{ }
     \end{subfigure}
     \caption{Numerical simulation of the reservoir computer performances  in presence of noise and neglecting noise effects. Error bars represent the standard deviation of the score over 100 different random partitions of test and train datasets.}
     \label{fig:score_vs_noise}
\end{figure}

The numerical simulations allow to study the behaviour of the reservoir computer in conditions difficult or impossible to reach experimentally. In Fig. \ref{fig:score_vs_noise} we compare a simulation with realistic noise (which is in agreement with measurements) and a simulation without noise.

\begin{figure}
     \centering
     \begin{subfigure}[b]{0.45\textwidth}
         \centering
         \includegraphics[width=\textwidth]{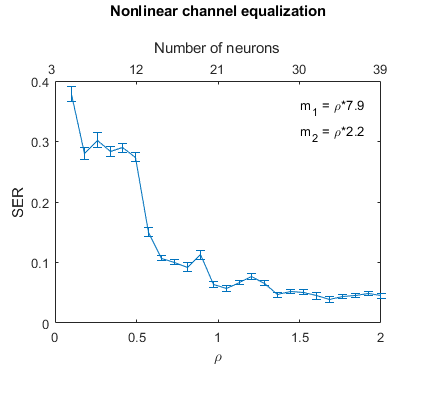}
         \caption{}
     \end{subfigure}
     \hfill
     \begin{subfigure}[b]{0.45\textwidth}
         \centering
         \includegraphics[width=\textwidth]{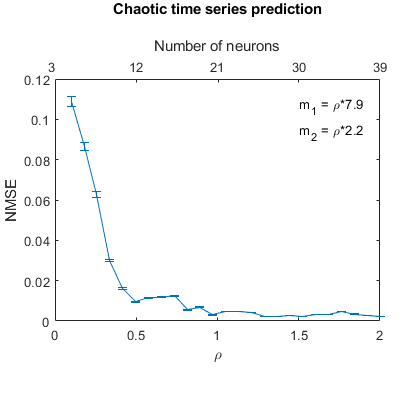}
         \caption{}
     \end{subfigure}
     \caption{Numerical simulation of the dependence of the reservoir computer performances on the strength of the phase modulation. Both $m_1$ (strength of the PM1 modulator) and $m_2$ (strength of the PM2 modulator) are scanned at the same time according to $m_1=\rho\cdot7.9$ and $m_2=\rho\cdot2.2$, with $\rho\in[0.1,2]$, and  $\rho=1$ corresponding to modulation strengths similar to the experimental ones. Top horizontal axes report the number of usable neurons (i.e.\ number of comb lines encoding a signal above the noise floor). Error bars represent the standard deviation of the score over 100 partitions of test and train datasets.  (a) Nonlinear channel equalization, SNR = $8$ dB (b) Chaotic time series prediction, shift = $-1$. 
     }
     \label{fig:score_vs_rf}
\end{figure}

In Fig.\ \ref{fig:score_vs_rf} we study the dependence of the performances on the strength of the phase modulation, varying the parameters $m_1$ and $m_2$.  
Smaller (larger) values of $m_1$ and $m_2$ correspond to a reservoir computer with less (more) neurons (i.e.\ comb lines). Indeed when $m_1$ and $m_2$ are small, there will be more very small amplitude comb lines which will be masked by noise. In Fig.\ \ref{fig:score_vs_rf} we also indicate the approximate number of comb lines that can be used, i.e.\ the effective number of neurons in the reservoir computer.
%Added the last sentences%

\newpage

\bibliography{bibl}{}
\bibliographystyle{ieeetr}

\end{document}